\documentclass[journal]{IEEEtran}

\usepackage{graphicx}
\usepackage{float}
\usepackage{mathrsfs}
\usepackage{amsfonts}
\usepackage{subeqn}

\usepackage{cite}

\usepackage{booktabs} 

\ifCLASSINFOpdf

\else

\fi

\begin{document}

\title{BER Analysis of Decision-Feedback Multiple Symbol Detection in Noncoherent MIMO Ultra-Wideband Systems}

\author{Taotao Wang, \emph{Student Member, IEEE,}, Tiejun Lv, \emph{Senior Member, IEEE,} Hui Gao, \emph{Member, IEEE}, Yueming Lu

\thanks{T. Wang is with the Department of Information Engineering, The Chinese University of Hong Kong, Hong Kong (email: wtt011@ie.cuhk.edu.hk). }
\thanks{T. Lv and Y. Lu are with the School of Information and Communication Engineering, Beijing University of Posts and Telecommunications, China (email: lvtiejun@bupt.edu.cn). }
\thanks{H. Gao is with Singapore University of Technology and Design, Singapore (email: hui\_gao@sutd.edu.sg). }
\thanks{T. Wang is supported by the Hong Kong PhD Fellowship Scheme. T. Lv and Y. Lu are supported by the National Natural Science Foundation of China (NSFC) (Grant No. 61271188) and the China 973 Program  (No. 2011CB302702).}
}
\markboth{This work has been accepted by IEEE Transactions on Vehicular Technology for publication}%
{Shell \MakeLowercase{\textit{et al.}}: Bare Demo of IEEEtran.cls
for Journals}

\maketitle

\begin{abstract}
In this paper, we investigate noncoherent multiple-input multiple-output (MIMO) ultra-wideband (UWB) systems where the signal is encoded by differential space-time block code (DSTBC).  DSTBC enables noncoherent MIMO UWB systems to achieve diversity gain.  However, the traditional noncoherent symbol-by-symbol differential detection (DD) for DSTBC-UWB suffers from performance degradation compared with the coherent detection.  We introduce a noncoherent multiple symbol detection (MSD) scheme to enhance the performance of DSTBC-UWB system.  Although the MSD scheme can boost the performance more as the observation window size gets to larger,  the complexity of the exhaustive search for MSD also exponentially increases in terms of the window size. To decrease the computational complexity, the concept of decision-feedback (DF) is introduced to MSD for DSTBC-UWB in this paper. The resultant DF-MSD yields reasonable complexity and also solid performance improvement. We provide  the bit error rate (BER) analysis for the proposed DF-MSD. Both theoretical analysis and simulation results validate the proposed scheme.
\end{abstract}


\IEEEpeerreviewmaketitle

\section{Introduction}
\IEEEPARstart{U}{ltra-wideband} (UWB) impulse radio communication, as a promising
candidate for location-aware indoor communications, wireless sensor
networks (WSN) and wireless personal area network (WPAN), has
attracted significant attention in both academia and industry in
recent years \cite{witrisal2009noncoherent,yang2004uwc}. Antenna array or known as
multiple-input multiple-output (MIMO) techniques, employing multiple
antennas at both link ends, is capable of improving system
performance by achieving spatial diversity. To exploit the
advantages of both UWB and MIMO systems, many researches have been done
on deploying the space-time coding (STC) scheme for MIMO UWB
transmissions \cite{yang2004ast, abou2007space}. STC schemes provide
diversity and coding gains for MIMO UWB systems and hence yield
remarkable performance improvement. However, the decoding operations
of these schemes require channel state information (CSI) and
implementations of Rake receiver. Since the UWB channel is
characterized by dense multipath,  stringent requirements on channel
estimation \cite{lottici2002channelestimation} and the number of
Rake fingers \cite{win1998ecu} make it difficult and costly to
realize the MIMO UWB system with Rake receiver.

In order to bypass the complicated treatments on UWB channel,
noncoherent UWB systems are proposed with good
performance-complexity tradeoff \cite{witrisal2009noncoherent}. The typical noncoherent UWB schemes are differential detection (DD) \cite{ho2002differential} and transmitted reference (TR) systems \cite{choi2002performance}, both employ analog autocorrelation receiver (AcR) that does not require channel estimation. Furthermore, noncoherent MIMO UWB systems are emerging recently \cite{abou2008simple, zhang2008dstbc}. The counterpart of DD in MIMO UWB is proposed in \cite{zhang2008dstbc}, wherein the signals are encoded by using differential space-time block code (DSTBC) \cite{hughes2000differential}. And a DD scheme employing MIMO AcR is further developed to exploit both the multipath diversity and spatial diversity. A performance gain, therefore, is obtained by MIMO DD scheme compared with single-input single-output (SISO) DD scheme in UWB systems.

Multiple symbol detection (MSD) is an effective means of improving
performance for noncoherent UWB systems. The theoretical framework
of MSD is the maximum-likelihood (ML) sequence detection, which is
firstly introduced to detect differential MPSK signal over additive
white Gaussian noise (AWGN) channel \cite{divsalar1990multiple}. MSD
is extended to MIMO system by \cite{gao2multiple}, where the author
uses MSD to detect DSTBC signal over a flat fading channel. Applying
MSD to SISO UWB system is considered in \cite{guo2006improved,
yang2008noncoherent} recently. Although MSD scheme can obtain a
solid performance improvement compared with symbol-by-symbol DD
scheme, the complexity of MSD exponentially increasing with
observation window size is intractable, especially in UWB systems
where high complexity receivers are not preferable.
Work \cite{lottici2008multiple} devises the application of sphere
decoding (SD) and Viterbi algorithm to fulfill a low complexity MSD for
differential SISO UWB system. In \cite{zhao2006decision},  the decisions on previous symbols are fed back to clean up the current noisy waveform template of UWB AcR. Then, this idea of decision-feedback (DF)  is employed to reduce the complexity of MSD for SISO UWB systems in \cite{qi2010fast, schenk2011decision}.

In this paper, we investigate noncoherent MIMO UWB systems where the UWB pulses are encoded by DSTBC as in \cite{hughes2000differential}.  The MSD can be employed to enhance the detection performance of DSTBC-UWB systems.  Nevertheless, the exhaustive search of the MSD scheme involves prohibitive computational complexity with large observation window size. In \cite{wang2011sphere}, SD is used to achieve practical implementation of MSD for DSTBC-UWB. Although SD scheme can facilitate the implementation of MSD,  its exponential expected complexity given fixed SNR \cite{jalden2005complexity} could be unaffordable when observation window size is rather large. This motivates us to propose a new DF based MSD scheme for DSTBC-UWB which is much simpler as well as more effective. With the concept of DF, we substitute the previous detected symbols into the MSD metric, thus reduce the search space. We still explore the correlations among multiple symbols to get more information assisting the detection of the current symbol.  We derive the theoretical lower bound of the bit error rate (BER) for the proposed DF-MSD scheme. It is shown in both theoretical analysis and computer simulations that the proposed DF-MSD scheme achieves a solid performance and outperforms DD scheme with slightly higher complexity. This paper is an extension of our conference paper \cite{wang2011noncoherent} where the BER performance analysis is lacking.

The rest of this paper is organized as follows. The system model of
DSTBC-UWB system and the MSD receiver scheme are described in Section II. Section III introduces the DF-MSD scheme and investigates the theoretical error performance of the DF-MSD. Section IV shows simulation results. Finally, conclusions are drawn in Section V.

\section{System Model}
\subsection{Transmit Signal Model}

A peer to peer MIMO UWB communication system includes a transmitter
equipped with $N_t$ antennas and a receiver equipped with $N_r$
antennas. Being consistent with \cite{zhang2008dstbc}, we focus on
the system with $N_t=2$ transmit antennas which is desirable in the
practical case. In the system where DSTBC scheme is employed to encode the
original information bits, the input bit sequence is divided into
blocks of $2$ bits; and then these bit blocks are fed into the encoder.
The encoder maps each $2$ bits into a space-time block codeword drawn from a codeword book $\Omega = \left\{
{{\bf{U}}^{\left( 1 \right)} ,{\bf{U}}^{\left( 2 \right)}
,{\bf{U}}^{\left( 3 \right)} ,{\bf{U}}^{\left( 4 \right)} }
\right\}$ where the codewords
\begin{equation}
\begin{array}{l}
 {{\bf{U}}^{\left( 1 \right)}} = \left( {\begin{array}{*{20}{c}}
   \;\;\;1 &  \;\;\;0\;  \\
    \;\;\;0 &  \;\;\;1 \; \\
\end{array}} \right), {{\bf{U}}^{\left( 2 \right)}} = \left( {\begin{array}{*{20}{c}}
    \;\;\;0 &  \;\;\;1\;  \\
   - 1 &  \;\;\;0 \; \\
\end{array}} \right),
 \\{{\bf{U}}^{\left( 3 \right)}} = \left( {\begin{array}{*{20}{c}}
   { - 1} &  \;\;\;0\;  \\
   \;\;\;0 & { - 1}\;  \\
\end{array}} \right), {{\bf{U}}^{\left( 4 \right)}} = \left( {\begin{array}{*{20}{c}}
    \;\;\;0 &  -1  \;\\
    \;\;\;1 &  \;\;\;0\;  \\
\end{array}} \right), \\
 \end{array}
\end{equation}
are constructed according to the property  ${\bf{U}}^{\left( m
\right)} {{\bf{U}}^{\left( m \right)}} ^{\rm{T}}  =
{{\bf{U}}^{\left( m \right)} }^{\rm{T}} {\bf{U}}^{\left( m \right)}
= {\bf{I}}$ for all $ {\bf{U}}^{\left( m \right)}  \in \Omega $
\cite{hughes2000differential}, ${\left(  \cdot \right)^{\rm{T}}}$ is
the transpose operator. When $N_t>2$,  we can also construct codewords
according to this property.  The proposed scheme, therefore, can be
extended to a system with larger $N_t$. The bits-to-codeword mapping
is Gray mapping such that $00$ $\to$ $ {\bf{U}}^{(1)}$, $01$ $\to$ $
{\bf{U}}^{(2)}$, $11$ $\to$ $ {\bf{U}}^{(3)}$ and $10$ $\to$ $
{\bf{U}}^{(4)}$. Then the information-bearing codewords are
differentially encoded to obtain the transmitted symbols:
\begin{equation}
{\bf{D}}_k  = {\bf{D}}_{k - 1} {\bf{U}}_k,\;\;\; k=1,2,\cdots
,\infty,
\end{equation}
where the $2 \times 2$ matrix ${\bf{D}}_k$ is the $k^{th}$
transmitted symbol which will be transmitted over $2$ transmit
antennas during $2$ frame durations, $ {\bf{U}}_k \in \Omega$ is the
$k^{th}$ information-bearing symbol (a codeword picked from $\Omega$). The initial
transmitted symbol ${\bf{D}}_0$ is the reference symbol which is set to as ${{\bf{D}}_0} = \left( {\begin{array}{*{20}{c}}
   \;1 & \;\;\;1  \\
   \;1 & { - 1}  \\
\end{array}} \right)$.
It is worth noting that ${\bf{D}}_{k}$ meets the orthogonal property ${\bf{D}}_k {\bf{D}}_{k} ^ {\rm{T}}  ={\bf{D}}_k ^{\rm{T}}{\bf{D}}_k
=2{\bf{I}}$, $\forall k$. The $p^{th}$ row and $n^{th}$ column entry of ${\bf{D}}_k$, denoted
by $d_{p,2k+n} \in \left\{ { \pm 1} \right\}$, is transmitted by
the $p^{th}$ transmit antenna during the $n^{th}$ frame duration of
the $k^{th}$ transmitted symbol, where $p,n = 1,2$. Hence, the
signal radiated by the $p^{th}$ transmit antenna is given by
\begin{equation}\label{transmit signal}
\begin{array}{l}
s_{p} \left( t \right) = \sqrt {\frac{E_b }{2}} \sum\limits_{k =
0}^\infty  {\sum\limits_{n = 1}^2 {d_{p,2k+n}\omega \left( {t -
\left( {n - 1} \right)T_f  - kT_s } \right)} } \\ \;\;\;\;\;\;\;\;\;=\sqrt {\frac{{{E_b}}}{2}} \sum\limits_{j =
1}^\infty {{d_{p,j}}\omega \left( {t - \left( {j - 1} \right){T_f}}
\right)},
\end{array}
\end{equation}
where $\omega \left( t \right)$ is the monocycle pulse waveform of
duration $T_\omega$ with normalized energy, $T_f$ is the frame
duration (pulse repetition interval), $T_s=2T_f$ is the duration of
one transmitted symbol, $E_b  $ is the energy used to transmit one
bit information and normalization factor $2$ insures the same
transmission power level as in the single-antenna case. Note that the
second equality of (\ref{transmit signal}), where a single time
index $j = 2k + n$ is introduced to replace the double time index
$\left( {k,n} \right)$, correspondingly, $d_{p,2k+n}$ is rewritten
as $d_{p,j}$. In the following, we will alternately use $d_{p,2k+n}$ and $d_{p,j}$ depending on which one is more convenient for description. Since $\omega \left( t \right)$ has a very short duration, $T_\omega$ is typically on the order of nanoseconds, the transmitted signal occupies a huge bandwidth. The frame duration $T_f$ is usually hundred or thousand times longer than $T_\omega$, resulting in a low duty transmission. Since the single user case is considered, the time-hopping (TH) codes or direct-sequence (DS)
codes for multiple access is eliminated for simplicity.

\subsection{Received Signal Model}
The channel impulse response (CIR) between the $p^{th}$  transmit
antenna and the  $q^{th}$ receive antenna is modeled as
\begin{equation}
h_{q,p} (t) = \mathop \sum \limits_{l = 0}^{L - 1} \alpha _l^{q,p}
\delta \left( {t - {\tau _l^{q,p} } } \right),
\end{equation}
where $\delta$ is the Dirac delta function, $L$ is the number of
resolvable multipath components (MPCs),  $\alpha _l^{q,p}$ and $\tau
_l^{q,p}$ are the gain and delay of each MPC, respectively. The
received signal at the $q^{th}$ receive antenna is given by
\begin{equation}
\begin{array}{l}
 {r_q}\left( t \right) = \sum\limits_{p = 1}^{{2}} {{s_p}\left( t \right)}  \otimes {h_{q,p}}\left( t \right) + {n_q}\left( t \right)\\ \;\;\;\;\;\;\;\;\;=\sqrt {\frac{{{E_b}}}{2}} \sum\limits_{p = 1}^2 {\sum\limits_{j = 1}^\infty  {{d_{p,j}}{g_{q,p}}\left( {t - \left( {j - 1} \right){T_f}} \right)} }  + {n_q}\left( t
 \right),
 \\
 \end{array}
\end{equation}
where $\otimes$ stands for convolution, $n_q\left( t \right)$
denotes zero mean AWGN process with two-sided power spectral density
${{N_0 } \mathord{\left/
 {\vphantom {{N_0 } 2}} \right.\kern-\nulldelimiterspace} 2}$, and $
{g_{q,p}\left( t \right)}$ is the overall channel response between
the $p^{th}$ transmit antenna and the $q^{th}$ receive antenna which
can be defined as
\begin{equation}
{g_{q,p}}\left( t \right) = \omega \left( t \right) \otimes
{h_{q,p}}\left( t \right)=\sum\nolimits_{l = 1}^L {\alpha
_l^{q,p}\omega \left( {t - \tau _l^{q,p}} \right)}.
\end{equation}
With $T_g$ denoting the maximum delay spread of all ${g_{q,p}}\left(
t \right)$, the inter-symbol interference (ISI) is avoided by letting
${T_f} \ge {T_g}$. For noncoherent reception, all ${g_{q,p}}\left( t
\right)$ are not known at the receiver.

\subsection{MSD Receiver for DSTBC-UWB}
In order to detect $M-1$ information-bering symbols jointly, the observation window length of MSD receiver is set to $M$ symbol durations, during which we assume that
the channel remains invariant. Since the UWB channel is quasi-static
in typical indoor environments \cite{channel}, this assumption is
justifiable.  For notational simplicity but without loss of generality, we
consider the case of recovering the first $M-1$ information-bering
symbols, which are given by a set
$\mathbb{U}=[{{\bf{U}}_1},{{\bf{U}}_2}, \cdots ,{{\bf{U}}_{M - 1}}]$,
from the observation $\left\{ {{r_q}(t)} \right\}_{q = 1}^{{N_r}}$
where $0 \le t \le M T_s$.  The detection strategy of the MSD for DSTBC-UWB system is
expressed as
\begin{equation}\label{detection strategy1}
\widehat{\mathbb{U}} = \mathop {\arg \max
}\limits_{\widetilde{\mathbb{U}} \in {\Omega ^{M - 1}}} \Lambda
\left( {\left\{ {{r_q}\left( t \right)} \right\}_{q =
1}^{{N_r}}\left| {\widetilde{\mathbb{U}}} \right.} \right)
\end{equation}
with
\begin{equation}\label{detection strategy2}
\begin{array}{l}
\Lambda \left( {\left\{ {{r_q}\left( t \right)} \right\}_{q =
1}^{{N_r}}\left| {\widetilde{\mathbb{U}}} \right.} \right) \\=
\sum\limits_{k = 1}^{M - 1} {\sum\limits_{y = 0}^{k - 1}
{{\rm{Tr}}\left( {\left( {\mathop \prod \limits_{v = y + 1}^k
{{\widetilde{\bf{U}}}_v}} \right){{ \left( {\sum\limits_{q = 1}^{{N_r}} {{\bf{R}}_{k,y}^q} } \right)
    }}} \right)} },
\end{array}
\end{equation}
where  ${\rm{Tr}}\left(  \cdot  \right)$ denotes the trace of a
matrix, $ \widetilde{\mathbb{U}}= [{\widetilde{\bf{U}}_1},{\widetilde{\bf{U}}_2}, \cdots ,{\widetilde{\bf{U}}_{M - 1}}]$ is the set of trail symbols, and ${\bf{R}}_{k,y}^q$ is a $2 \times 2$ matrix whose entries
are the correlations between the $k^{th}$  symbol
waveform and the $y^{th}$  symbol waveform from the $q^{th}$  receive antenna. Upon defining the
vector of the time-shifted received continuous time waveforms at the $q^{th}$ antenna as ${\bf{r}}_k^q \left(
t \right) = \left[ {\begin{array}{*{20}c}
   {r_q \left( {t + 2kT_f } \right)} & {r_q \left( {t + (2k+1)T_f } \right)}  \\
\end{array}} \right]$, ${\bf{R}}_{k,y}^{q}$ can be expressed as
\begin{equation}\label{correlation}
{\bf{R}}_{k,y}^q = \int_0^{{T_i}} {{{\left( {{\bf{r}}_k^q\left( t \right)} \right)}^{\rm{T}}}} {\bf{r}}_y^q\left( t \right)dt,
\end{equation}
where $T_i$ is the integration interval. Some remarks are given below.

(i) The detection metric of MSD (\ref{detection strategy2}) can be derived by using generalized likelihood ratio testing (GLRT) approach.  We refer the interested readers to \cite{wang2011noncoherent} for details on the  derivation.  When $M=2$, (\ref{detection strategy2}) degenerates into DD scheme for DSTBC-UWB system which also appears in \cite{zhang2008dstbc}.

(ii) The MSD scheme requires the computation of correlations between different segments of the received signal (as in (\ref{correlation})). This can be implemented by using analog delay line, multiplier and integrator which avoids analog to digital converter with ultra high sampling rate. The computed correlations will be fed into a digital processor which fulfills the search task of (\ref{detection strategy1}) . We illustrate the schematic structure of MSD receiver in Fig.1 with $M=3$ as an example.

(iii) As $M$ increases, the MSD scheme can offer an additional detection gain compared with DD. However, the detection strategy (\ref{detection strategy1}) is a ML sequence
detection that has an exhaustive search process, and the size of
search space grows exponentially with $M-1$. When $M$ gets large,
the computational complexity becomes intolerable.  The complexity of exhaustive search for MSD has been reduced in \cite{wang2011sphere} by SD. In next section, we apply the DF mechanism to further reduce the computational complexity of MSD for DSTBC-UWB.

\begin{figure*}[!t]
\normalsize
\includegraphics[width=5.5in]{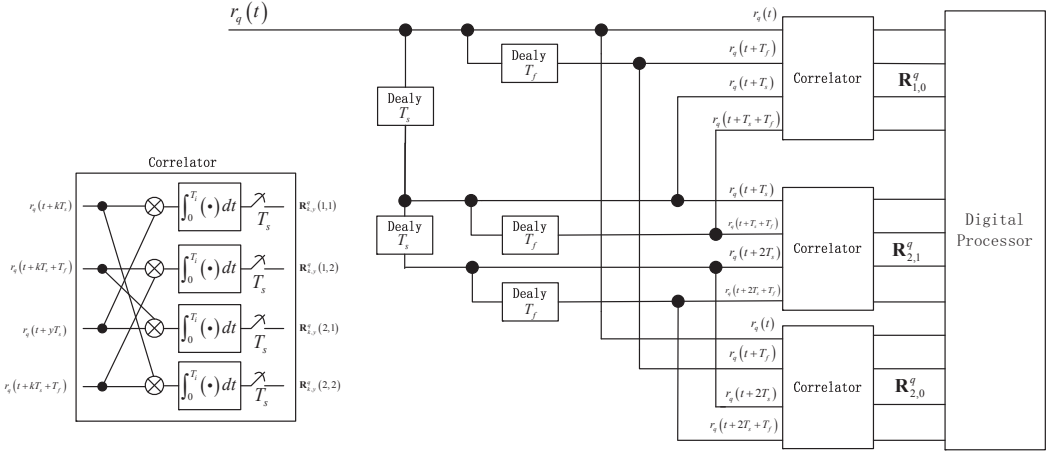}
\caption{Schematic structure of MSD receiver with $M=3$ as an example.} \label{ACR}
\end{figure*}

\section{DF based MSD for DSTBC-UWB}

\subsection{DF-MSD}
In this section, the DF-MSD with lower complexity than
the original MSD in  (\ref{detection strategy1}) is proposed to reduce the computational complexity. The notion of DF-MSD is to detect the $k^{th}$
information-bearing symbol ${{\bf{U}}_k}$ by substituting the
estimate of previous $M-2$ symbols, ${\widehat{\bf{U}}_{k - M +
2}},{\widehat{\bf{U}}_{k - M + 1}}, \cdots ,{\widehat{\bf{U}}_{k -
1}}$, into the original MSD metric (\ref{detection strategy1}). The
DF-MSD scheme for DSTBC-UWB is formalized as
\begin{equation}\label{DF1}
{\widehat{\bf{U}}_k} = \arg \mathop {\max }\limits_{{{\widetilde{\bf{U}}}_k \in \Omega  } } \Lambda \left( {\left\{ {{r_q}\left( t \right)} \right\}_{q = 1}^{{N_r}}\left| {{{\widetilde{\bf{U}}}_k}} \right.} \right)
\end{equation}
with
\begin{equation}\label{DF2}
\begin{array}{l}
\Lambda \left( {\left\{ {{r_q}\left( t \right)} \right\}_{q = 1}^{{N_r}}\left| {{{\widetilde{\bf{U}}}_k}} \right.} \right) \\= \begin{array}{*{20}{c}}
   {\sum\limits_{y = k - M + 1}^{k - 1} {{\rm{Tr}}\left( {\left( {\prod\limits_{v = y + 1}^{k - 1} {{{\widehat{\bf{U}}}_v}} } \right){{\widetilde{\bf{U}}}_k}{{\left( {\sum\limits_{q = 1}^{{N_r}} {{\bf{R}}_{k,y}^q} } \right)}}} \right)} .}  \\
\end{array}
\end{array}
\end{equation}
The observation window size of DF-MSD still preserves $M$, however,
the observation window slides one symbol duration down each time.
Since one codeword is detected each time, the size of search space
is reduced from ${\left| \Omega  \right|^{M - 1}}$ to $\left| \Omega  \right|$.  And, no signal correlations $\left\{ {{\bf{R}}_{k',y}^q} \right\}$, $\forall k' \ne k$ required to be computed, we just need to compute $\left\{ {{\bf{R}}_{k,y}^q} \right\}_{y = k - M + 1}^{k - 1}$ when detecting the $k^{th}$ symbol.  Thus, the DF-MSD has lower complexity than original MSD.

\subsection{Error Performance Analysis of DF-MSD}
In this subsection, the error performance of the proposed DSTBC-UWB system with DF-MSD receiver is evaluated. Since the errors among the $M - 2$ previous detected information-bearing codewords ${\widehat {\bf{U}}_{k - M + 2}},{\widehat {\bf{U}}_{k - M + 1}}, \cdots ,{\widehat {\bf{U}}_{k - 1}}$ have impact on the detection of the current codeword ${\widehat {\bf{U}}_k}$, an accurate expression for the error performance is intractable to derive. We turn to analyze the genie-aided DF-MSD receiver case \cite{schober1999decision} where all the $M - 2$ feedback codewords are assumed to be correct, i.e., ${\widehat {\bf{U}}_y} = {{\bf{U}}_y}$ for $y = k - M + 1, \cdots ,k - 1$. Its error performance is regarded as the lower bound of the real DF-MSD receiver.

Firstly, we investigate the details of ${\bf{R}}_{k,y}^q$ which used in (\ref{DF2}) for detection. The $2 \times 2$ matrix ${\bf{R}}_{k,y}^q$ is defined in (\ref{correlation}), and its ${u^{th}}$ row and $z^{th}$  column entry is computed as
\begin{equation}\label{corr}
\begin{array}{l}
R_{k,y}^q\left( {u,z} \right) = \int_0^{{T_i}} {{r^q}\left( {t + 2k{T_f} + \left( {u - 1} \right){T_f}} \right)} \\\;\;\;\;\;\;\;\;\;\;\;\;\;\;\;\;\;\;\;\;\;\;\;\;\;\;\;\;\;\;\;\;\;\;\;\;\;\;\;\;\; \times {r^q}\left( {t + 2y{T_f} + \left( {z - 1} \right){T_f}} \right)dt\\
 \;\;\;\;\;\;\;\;\; \;\;\;\;\;\;\;= \int_0^{{T_f}} {\left( {\sqrt {\frac{{{E_b}}}{2}} \sum\limits_{p = 1}^2 {{d_{p,2k + u}}{g_{q,p}}\left( t \right) + n_{k,u}^q\left( t \right)} } \right)} \\\;\;\;\;\;\;\;\;\;\;\;\;\;\;\;\;\;\;\;\;\;\;\;\; \times \left( {\sqrt {\frac{{{E_b}}}{2}} \sum\limits_{p = 1}^2 {{d_{p,2y + z}}{g_{q,p}}\left( t \right) + n_{y,z}^q\left( t \right)} } \right)dt,
\end{array}
\end{equation}
where
\begin{equation}
\begin{array}{l}
n_{k,u}^q\left( t \right) = {n_q}\left( {t + 2k{T_f} + \left( {u - 1} \right){T_f}} \right),\\ n_{y,z}^q\left( t \right) = {n_q}\left( {t + 2y{T_f} + \left( {z - 1} \right){T_f}} \right),
\end{array}
\end{equation}
are the time-shifted noises. Further expanding (\ref{corr}), ${{R}}_{k,y}^q$ is divided into a signal component and a noise component:
\begin{equation}
R_{k,y}^q\left( {u,z} \right) = S_{k,y}^q\left( {u,z} \right) + N_{k,y}^q\left( {u,z} \right),
\end{equation}
where the signal component is
\begin{equation}\label{signal component}
\begin{array}{l}
S_{k,y}^q\left( {u,z} \right) \\= \sum\limits_{p' = 1}^2 {\sum\limits_{p'' = 1}^2 {{d_{p',2k + u}}{d_{p'',2y + z}}} } \frac{{{E_b}}}{2}\int_0^{{T_i}} {{g_{q,p'}}\left( t \right){g_{q,p''}}\left( t \right)} dt,
\end{array}
\end{equation}
the noise component is
\begin{equation}\label{noise component}
N_{k,y}^q\left( {u,z} \right) = {A_1} + {A_2} + {A_3}
\end{equation}
with
\begin{equation}\label{A3}
\begin{array}{l}
{A_1} = \sum\limits_{p = 1}^2 {{d_{p,2k + u}}\sqrt {\frac{{{E_b}}}{2}} \int_0^{{T_i}} {{g_{q,p}}\left( t \right)n_{y,z}^q\left( t \right)dt} }, \\ {A_2} = \sum\limits_{p = 1}^2 {{d_{p,2y + z}}\sqrt {\frac{{{E_b}}}{2}} \int_0^{{T_i}} {{g_{q,p}}\left( t \right)n_{k,u}^q\left( t \right)dt} }, \\ {A_3} = \int_0^{{T_i}} {n_{k,u}^q\left( t \right)n_{y,z}^q\left( t \right)dt}.
\end{array}
\end{equation}
Now, ${\bf{R}}_{k,y}^q$ can be rewritten as a matrix form
\begin{equation}
{\bf{R}}_{k,y}^q = {\bf{S}}_{k,y}^q + {\bf{N}}_{k,y}^q,
\end{equation}
where the entries of matrixes
\begin{equation}
\begin{array}{l}
{\bf{S}}_{k,y}^q = \left[ {\begin{array}{*{20}{c}}
{S_{k,y}^q\left( {1,1} \right)}&{S_{k,y}^q\left( {1,2} \right)}\\
{S_{k,y}^q\left( {2,1} \right)}&{S_{k,y}^q\left( {2,2} \right)}
\end{array}} \right],\\
{\bf{N}}_{k,y}^q = \left[ {\begin{array}{*{20}{c}}
{N_{k,y}^q\left( {1,1} \right)}&{N_{k,y}^q\left( {1,2} \right)}\\
{N_{k,y}^q\left( {2,1} \right)}&{N_{k,y}^q\left( {2,2} \right)}
\end{array}} \right],
\end{array}
\end{equation}
are defined in (\ref{signal component}) and (\ref{noise component}).

Then, by assuming the $k^{th}$ transmitted codeword ${{\bf{U}}_k}$ is ${{\bf{U}}^{\left( 1 \right)}}$, we analyze the probability of the correct detection:
\begin{equation}\label{Pc}
{P_c} = P\left( {{\Psi _1} > {\Psi _2},{\Psi _1} > {\Psi _3},{\Psi _1} > {\Psi _4}} \right),
\end{equation}
where ${\Psi _m}$ is the detection metric for the trial codeword ${{\bf{U}}^{\left( m \right)}}$ and it is defined as
\begin{equation}
{\Psi _m} = \Lambda \left( {\left\{ {{r_q}\left( t \right)} \right\}_{q = 1}^{{N_r}}\left| {{{\widetilde{\bf{U}}}_k} = {{\bf{U}}^{\left( m \right)}}} \right.} \right),
\end{equation}
for each $m$. Using the computation about ${\bf{R}}_{k,y}^q$ and the assumption ${{\bf{U}}_k} = {{\bf{U}}^{\left( 1 \right)}}$, ${\Psi _1}$, ${\Psi _2}$, ${\Psi _3}$, ${\Psi _4}$ are given by
\begin{subequations}
\begin{equation}
\begin{array}{l}
{\Psi _1} = 2\left( {M - 1} \right)\frac{{{E_b}}}{2}\sum\limits_{q = 1}^{{N_r}} {\left( {{\varepsilon _{q,1}} + {\varepsilon _{q,2}}} \right)} \\\;\;\;\;\;\;\;\;\;\;\;\;\;\;\;\;\;\;\;\;\;\;+ \sum\limits_{q = 1}^{{N_r}} {\sum\limits_{y = k - M + 1}^{k - 1} {\left( {N_{k,y}^q\left( {1,1} \right) + N_{k,y}^q\left( {2,2} \right)} \right)} },
\end{array}
\end{equation}
\begin{equation}
{\Psi _2} = \sum\limits_{q = 1}^{{N_r}} {\sum\limits_{y = k - M + 1}^{k - 1} {\left( {N_{k,y}^q\left( {2,1} \right) + N_{k,y}^q\left( {1,2} \right)} \right)} },
\end{equation}
\begin{equation}\label{phi3}
{\Psi _3} =  - {\Psi _1},
\end{equation}
\begin{equation}\label{phi4}
{\Psi _4} =  - {\Psi _2},
\end{equation}
\end{subequations}
where ${\varepsilon _{q,p}} = {\int_0^{{T_i}} {\left( {{g_{q,p}}\left( t \right)} \right)} ^2}d$ is the captured energy of the channel between $p^{th}$ transmit antenna and the $q^{th}$  receive antenna. Substituting (\ref{phi3}) and (\ref{phi4})  in (\ref{Pc}), ${P_c}$ is expressed as
\begin{equation}
\begin{array}{l}
{P_c} = P\left( {{\Psi _1} > {\Psi _2},{\Psi _1} >  - {\Psi _1},{\Psi _1} >  - {\Psi _2}} \right) \\ \;\;\;\;\;= P\left( {{\Psi _1} > \left| {{\Psi _2}} \right|,{\Psi _1} > 0} \right).
\end{array}
\end{equation}
In order to evaluate ${P_c}$, the conditional correct probability ${P_{c\left| h \right.}}$ given the channel realizations $\left\{ {{h_{q,p}}\left( t \right)} \right\}$ is our target. Conditioning on the channels, ${A_1}$, ${A_2}$ are strictly Gaussian random variables; ${A_3}$ is approximately Gaussian \cite{quek2005analysis}. These conditional Gaussian random variables are also mutually independent \cite{quek2005analysis}. Therefore, the conditional probability density functions of Gaussian random variables ${\Psi _1}$ and ${\Psi _2}$ are
\begin{subequations}\label{PP}
\begin{equation}\label{fphi1}
{f_{{\Psi _1}\left| h \right.}}\left( {{\psi _1}} \right) = \frac{1}{{\sqrt {2\pi {\sigma ^2}} }}\exp \left( {\frac{{ - {{\left( {{\psi _1} - S} \right)}^2}}}{{2{\sigma ^2}}}} \right),
\end{equation}
\begin{equation}\label{fphi2}
{f_{{\Psi _2}\left| h \right.}}\left( {{\psi _2}} \right) = \frac{1}{{\sqrt {2\pi {\sigma ^2}} }}\exp \left( {\frac{{ - {{\left( {{\psi _2}} \right)}^2}}}{{2{\sigma ^2}}}} \right),
\end{equation}
\end{subequations}
with the condition mean of ${\Psi _1}$
\begin{equation}\label{SS}
S = {\rm E}\left( {{\Psi _1}\left| h \right.} \right) = {E_b}\left( {M - 1} \right)\varepsilon,
\end{equation}
and the conditional variance of ${\Psi _1}$  and ${\Psi _2}$
\begin{equation}\label{AA}
\begin{array}{l}
{\sigma ^2} = {\rm E}\left( {\Psi _2^2\left| h \right.} \right) = {\rm E}\left( {{{\left( {{\Psi _1} - S} \right)}^2}\left| h \right.} \right) \\ \;\;\;\;\;= \left( {M - 1} \right)\left( {{E_b}{N_0}\varepsilon  + {N_0}^2W{T_i}} \right),
\end{array}
\end{equation}
where $\varepsilon  = {N_r}\sum\nolimits_{q = 1}^{{N_r}} {\left( {{\varepsilon _{q,1}} + {\varepsilon _{q,2}}} \right)}$ is the total captured channel energy, $W$ is the bandwidth of the receiver band-pass filter. The detailed derivations of (\ref{PP})-(\ref{AA}) are given in Appendix. Using (\ref{fphi1}) and (\ref{fphi2}), the conditional correct probability ${P_{c\left| h \right.}}$ is given by
\begin{equation}
\begin{array}{l}
{P_{c\left| h \right.}} = P\left( {{\Psi _1} > \left| {{\Psi _2}} \right|,{\Psi _1} > 0\left| h \right.} \right) \\\;\;\;\;\;\;\;\;= \int_0^\infty  {P\left( {\left| {{\Psi _2}} \right| < {\Psi _1}\left| {{\Psi _1} > 0,h} \right.} \right)} {f_{{\Psi _1}\left| h \right.}}\left( {{\psi _1}} \right)d{\psi _1} \\
\\ \;\;\;\;\;\;\;\; = \int_0^\infty  {\left( {\int_{ - {\psi _1}}^{{\psi _1}} {{f_{{\Psi _2}\left| h \right.}}\left( {{\psi _2}} \right)d{\psi _2}} } \right)} {f_{{\Psi _1}\left| h \right.}}\left( {{\psi _1}} \right)d{\psi _1} \\\;\;\;\;\;\;\;\;= \int_0^\infty  {{\rm{erf}}\left( {\frac{{{\psi _1}}}{{\sqrt 2 \sigma }}} \right)\frac{1}{{\sqrt {2\pi {\sigma ^2}} }}\exp \left( {\frac{{ - {{\left( {{\psi _1} - S} \right)}^2}}}{{2{\sigma ^2}}}} \right)} d{\psi _1}.
\end{array}
\end{equation}
The conditional codeword error probability is ${P_{e\left| h \right.}} = 1 - {P_{c\left| h \right.}}$. Since the Gray mapping is used, the conditional BER is
\begin{equation}
{P_{ber\left| h \right.}} = \frac{1}{2}\left( {1 - {P_{c\left| h \right.}}} \right).
\end{equation}
Finally, the unconditional BER can be obtained by averaging over the channel
\begin{equation}\label{BER}
{P_{ber}} = \int {{P_{ber\left| h \right.}}P\left( h \right)dh},
\end{equation}
which is realized by numerical integration. It is noted that ${P_{ber}}$ is the BER of the genie-aided DF-MSD receiver which is regarded as the lower bound of our DF-MSD receiver.


\section {Numerical Experiments}


In this section, numerical experiments are conducted to validate the proposed schemes. In all experiments, the channel are generated according to IEEE 802.15.3a CM2 model \cite{channel}. The used impulse shape is the second derivative of a Gaussian function. The duration of $\omega(t)$ is set as $T_{\omega}=0.5$ ns. We assume there is no ISI in all systems, thus, the frame duration $T_f$ is set to $100$ ns which is larger than the maximum excess delay of the channel. The integration interval is $T_i=20$ ns. In all cases, the antenna setting is $N_t=2$, $N_r=1$.

%

\begin{figure}[!h]
\centering
\includegraphics[width=3in]{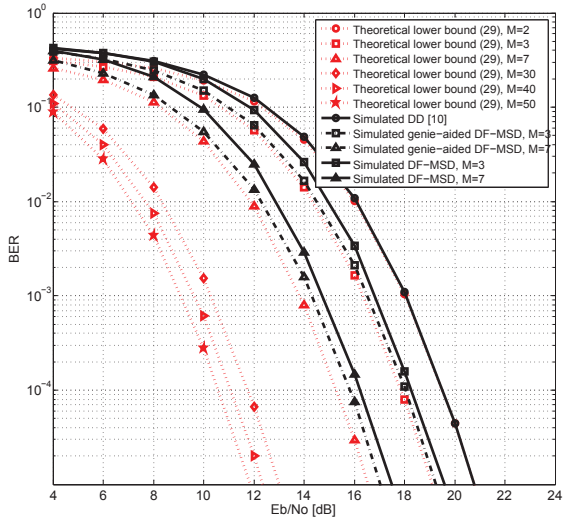}
\caption{The theoretical BER and simulated BER of DF-MSD.} \label{fig_sim1}
\end{figure}

\textbf{TEST 1: } We present numerical evaluation and simulation to verify the theoretical analysis. The theoretical BER for the genie-aided DF-MSD receiver (\ref{BER}) is computed by numerical integration. The BER performances of the genie-aided DF-MSD receiver and the real DF-MSD receiver which is described by (\ref{DF1}) are also given by computer simulations. The simulated BER curves together with the theoretical ones are shown in Fig. \ref{fig_sim1} for various $M$. It can be observed that the BER of the genie-aided DF-MSD receiver is less than that of real DF-MSD receiver. This performance gap is due to the error propagation from the feedback codewords in real DF-MSD receiver. For the theoretical BER of genie-aided DF-MSD, it does also not totally coincide with the simulated one. This inaccuracy is caused by the approximation of $A_3$ in (\ref{A3}) as Gaussian.  The theoretical BER can provide good performance approximation since it is a lower bound and the bound is somewhat tight. When $M$  gets too large, the complexity of the proposed DF-MSD becomes very high. It is impossible to simulate such system. But, we can evaluate the theoretical lower bound to predict the BER performance of the proposed DF-MSD with larger $M$.

\begin{figure}[!h]
\centering
\includegraphics[width=3.1in]{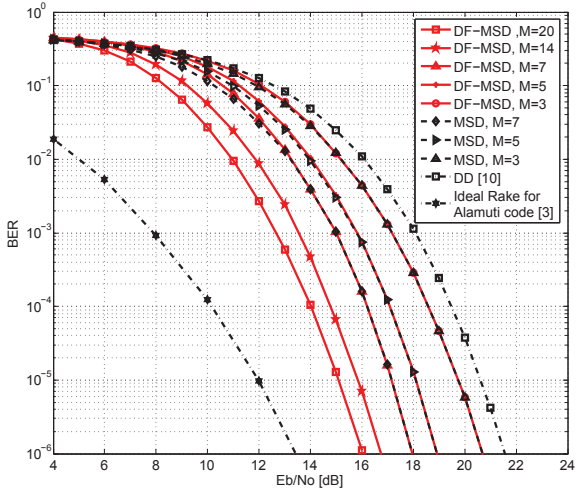}
\caption{The BER comparisons of DF-MSD for DSTBC-UWB
system between other schemes.} \label{fig_sim2}
\end{figure}

\textbf{TEST 2:}  In this case, we compare the BER performance of
DSTBC-UWB systems with DF-MSD described by (\ref{DF1})  with other schemes. The BER performances
of ideal Rake reception for Alamuti space-time coded MIMO UWB system
\cite{yang2004ast}, DD scheme for DSTBC-UWB
\cite{zhang2008dstbc} system and MSD for DSTBC-UWB system described by (\ref{detection strategy1})  are also evaluated and presented  as the performance benchmarks. The results are shown in Fig. \ref{fig_sim2}.  First, we can observe the clear performance benefit brought by MSD compared with DD scheme. Then,  we can find that DF-MSD for DSTBC-UWB almost have the same BER with MSD except that DF-MSD slightly underperforms
MSD when ${{{E_b}} \mathord{\left/
 {\vphantom {{{E_b}} {{N_0}}}} \right.
 \kern-\nulldelimiterspace} {{N_0}}}$ is small. This slight performance
gap is due to the error propagation effect of DF-MSD at low
${{{E_b}} \mathord{\left/
 {\vphantom {{{E_b}} {{N_0}}}} \right.
 \kern-\nulldelimiterspace} {{N_0}}}$. Finally, since the search space of
DF-MSD will not increase with observation window size $M$, we can
increase $M$ to enable the BER get close to ideal Rake reception.
For example, the performance gap between the proposed DF-MSD for DSTBC-UWB with
$M=20$ and ideal Rake reception for Alamuti space-time coded
MIMO-UWB is within $3$ dB at BER=$10^{-6}$.

%


\section{Conclusion}
In this paper, we consider a differential space-time block coded
MIMO UWB system. We employ MSD scheme to improve detection performance. Moreover, to reduce the complexity of MSD and make its realization more practical, we introduce the DF-MSD scheme. Since
the search space of DF-MSD will not increase with observation window
size $M$ but more information can be used to enhance detection
performance by increasing $M$, the BER is improved by the DF-MSD
with moderate detection complexity. We give BER analysis of the genie-aided  DF-MSD scheme, which can be served as the theoretical lower bound for the BER of real DF-MSD and also can be used to guide the soft detector design in the future work.  The simulation results coincide with our theoretical analysis and validate the proposed scheme.

\appendix
This appendix gives the derivations of ${f_{{\Psi _1}\left| h \right.}}$ and ${f_{{\Psi _2}\left| h \right.}}$.  Obviously, the conditional means of ${\Psi _1}$, ${\Psi _2}$ are
\begin{equation}
\begin{array}{l}
{\rm E}\left( {{\Psi _1}\left| h \right.} \right) = \left( {M - 1} \right){E_b}\varepsilon, \\  {\rm E}\left( {{\Psi _2}\left| h \right.} \right) = 0.
\end{array}
\end{equation}
Given the channel realizations, ${A_1}$, ${A_2}$, ${A_3}$ are independent zero mean Gaussian random variables, and the conditional variances of them are \cite{quek2005analysis}
\begin{equation}
\begin{array}{l}
{\rm E}\left( {{A_1}^2\left| h \right.} \right) = \frac{{{E_b}}}{2}\frac{{{N_0}}}{2}\left( {{\varepsilon _{q,1}} + {\varepsilon _{q,2}}} \right), \\
{\rm E}\left( {{A_2}^2\left| h \right.} \right) = \frac{{{E_b}}}{2}\frac{{{N_0}}}{2}\left( {{\varepsilon _{q,1}} + {\varepsilon _{q,2}}} \right), \\
{\rm E}\left( {{A_3}^2\left| h \right.} \right) = \frac{{{N_0}^2}}{2}W{T_i}.
\end{array}
\end{equation}
Then, the conditional variance of $N_{k,y}^q\left( {u,z} \right)$ is
\begin{equation}
E\left( {{{\left( {N_{k,y}^q\left( {u,z} \right)} \right)}^2}\left| h \right.} \right) = \frac{{{E_b}}}{2}{N_0}\left( {{\varepsilon _{q,1}} + {\varepsilon _{q,2}}} \right) + \frac{{{N_0}^2}}{2}W{T_i}.
\end{equation}
Therefore, the conditional variances of ${\Psi _1}$, ${\Psi _2}$ are
\begin{equation}
\begin{array}{l}
{\rm E}\left( {{{\left( {{\Psi _1} - S} \right)}^2}\left| h \right.} \right) = \left( {M - 1} \right)\left( {{E_b}{N_0}\varepsilon  + {N_0}^2W{T_i}} \right),\\ {\rm E}\left( {\Psi _2^2\left| h \right.} \right) = \left( {M - 1} \right)\left( {{E_b}{N_0}\varepsilon  + {N_0}^2W{T_i}} \right).
\end{array}
\end{equation}
After the conditional means and variances of ${\Psi _1}$, ${\Psi _2}$ are calculated, the probability density functions of them are given by (\ref{fphi1}), (\ref{fphi2}).

\ifCLASSOPTIONcaptionsoff
  \newpage
\fi

\bibliographystyle{IEEEtran}
\bibliography{symbollevelcombining}


\end{document}